\documentclass[aps,prf]{revtex4}{

\usepackage{graphicx}

\usepackage{amsmath,amsfonts,bm, color,amssymb,float}
\usepackage{graphicx,epsfig,overpic,hyperref,ulem,caption,subcaption}
\usepackage{siunitx}

\begin{document}

\title{Turbulent-like flows in quasi two-dimensional dense suspensions of motile colloids}
\author{Rui Luo\textsuperscript{1}, Alexey Snezhko\textsuperscript{2}, and Petia M. Vlahovska\textsuperscript{1}}

\affiliation{(1) Engineering Sciences and Applied Mathematics, Northwestern University, Evanston, IL 60208, USA \\
(2) Materials Science Division, Argonne National Laboratory, 9700 South Cass Avenue,
Lemont, Illinois 60439, USA.}

\date{}

\begin{abstract}
Dense bacterial suspensions exhibit turbulent-like flows at low Reynolds numbers, driven by the activity of the microswimmers. In this study, we develop a model system to examine these dynamics using motile colloids that mimic bacterial locomotion. The colloids are powered by the Quincke instability, which causes them to spontaneously roll in a random-walk pattern when exposed to a square-wave electric field. We experimentally investigate the flow dynamics in dense suspensions of these Quincke random walkers under quasi two-dimensional conditions, where the particle size is comparable to the gap between the electrodes. 
Our results reveal an energy spectrum scaling at high wavenumbers as $ \sim k^{-4}$, which holds across a broad range of activity levels -- controlled by the field strength -- and particle concentrations. We observe that velocity time correlations decay within a single period of the square-wave field, yet an anti-correlation appears between successive field applications, indicative of a dynamic structural memory of the ensemble.

\end{abstract}

\date{ \today}

\maketitle

Swimming bacteria self-organize into macroscopic patterns such as swarms and dynamic clusters \cite{Copeland:2009, zhang:2009, zhang:2010, chen:2012, chen:2015,Beer:2020,Aranson:2022}. At high concentrations, turbulent-like motion emerges  characterized by erratic flows and transient vortices \cite{Dombrowski:2004, Sokolov:2007, wensink2012, Dunkel:2013, Gachelin_2014,Zhou:2014,peng2021}. 
Unlike the classical hydrodynamic turbulence, which is a high-Reynolds number, inertia-dominated phenomenon, bacterial turbulence occurs at very low Reynolds numbers, where inertia is negligible, and is driven by the active motion of the microswimmers \cite{alert2022}. 
The energy spectrum in classical turbulence in two dimensions typically follows a power-law dependence on the wavenumber, $E \sim k^{-5/3}$, whereas for bacterial turbulence different power laws have  been reported \cite{wensink2012,Liu-Cheng:2021SM,wei2024}. 
\begin{figure*}
	\includegraphics[width=\linewidth]{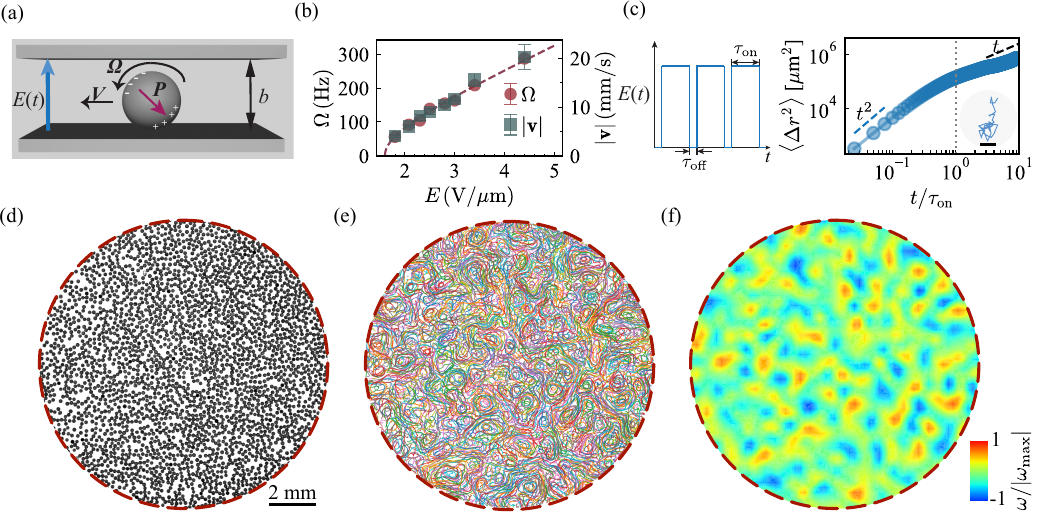}
	\caption{ \footnotesize { 
{\bf{Turbulent-like flows in suspension of Quincke Random Walkers:}}
(a) In a uniform direct current (DC) electric field \( E \), free charges  accumulate at the particle surface. 
Above the threshold for the  Quincke instability, 
the induced electric dipole tilts and
generates a net torque, causing the sphere to roll in a randomly chosen direction in the plane perpendicular to the applied field direction.  
(b) The rotation rate as well as the rolling speed, i.e., activity, of an individual Quincke roller increases with field strength.
(c) Temporal modulation of the electric field in the form of a square-wave causes the particle to undergo a random walk \cite{karani2019}, if
\( \tau_{\text{off}} \) is much longer than the ``particle memory'' time \( \tau_{\text{MW}} \) (the time required for complete depolarization). The mean-squared displacement ($\langle \Delta r^2 \rangle$) of a single Quincke random walker is shown for \( E = 4.4 \, \mathrm{V/\mu m} \), \( \tau_{\mathrm{on}} = 40 \, \mathrm{ms} \), and \( \tau_{\mathrm{off}} = 20 \, \mathrm{ms} \), with the corresponding trajectory displayed in the inset. Scale bar: \( 1 \, \mathrm{mm} \).
(d) Snapshot of the 
colloidal suspension  within the field of view of the microscope. (e) Particles trajectories, tracked using Particle Tracking Velocimetry (PTV), during one period \( \tau_{\text{on}} \).  (f) Vorticity computed based on the velocity field from Particle Imaging Velocimetry (PIV). The red dashed line indicates the field of view, not a physical boundary. The experiment is conducted at $E = 7.4\, \mathrm{V/um}$, particle area fraction $\phi = 0.51$.  
}}
	\label{fig:setup}
\end{figure*}

Active turbulence
has been observed in various living fluids such as sperm \cite{creppy:2015}, microtubule-kinesin bundles (active nematics) \cite{Dogic:2012, Opathalage:2019, Sagues:2019}, and cell tissues \cite{Blanch:2019, lin2021}. The phenomenon generated a lot of theoretical effort \cite{SR:2002, Saintillan07, Saintillan:2011, Saintillan:2013, wensink2012, Dunkel:2013, Dunkel:2013b, bratanov2015, giomi2015, James:2018, Doostmohammadi:2018, Linkmann:2019,Skultety-Morozov:2020,Shaebani:2020review,qi2022,w.zantop2022,Keta:2024,yang2024shaping}, to decipher how ``activity engenders turbulence'' \cite{Fraden:2019comm}. Recently, a universal scaling of $E(k) \sim k^{-4}$ has been predicted and validated for two-dimensional active nematics systems \cite{martinez-prat2021}, and similar scaling has been observed in studies of bacterial turbulence transitioning from two-dimensional to three-dimensional flow, where such scaling is absent in purely two-dimensional conditions \cite{wei2024}. However, experimental research in bacterial  turbulence lags the theoretical advances due to the challenges to have well defined and controllable conditions e.g., particle density, speed (i.e., activity) and locomotion type, when working with living microswimmers. Synthetic externally driven colloids offer an  alternative platform for studying active turbulence~\cite{nishiguchi2015,kokot2017active,mecke2023simultaneous}. 

A promising model system to emulate bacterial flows is based on the Quincke rollers \cite{bricard2013,Driscoll:2019,BOYMELGREEN:2022,Diwakar:2022,Bishop:2023}.
Their motility is due to an electrohydrodynamic instability which gives rise to a constant electric torque on  the particles   \cite{quincke1896} causing them to roll, if the particles are initially resting on the electrode. 
Quincke rollers exhibit a wide range of intricate  collective behaviors
\cite{bricard2013, bricard2015,chardac2021emergence,zhang2020reconfigurable,zhang2022,jorge2024active}. Temporal modulation of the rollers' activity through application of a square-wave electric field
modifies the rollers persistence length resulting in even richer collective dynamics \cite{karani2019,zhang2021persistence,zhang2022natphys}. By adjusting the durations of the pulses, $\tau_{\text{on}}$, and their spacing, $\tau_{\text{off}}$, the trajectory of an isolated Quincke roller 
becomes a random walk, and the particle locomotion can be tuned to emulate various bacterial motility patterns \cite{karani2019}.  
Populations of Quincke  random walkers exhibit behaviors reminiscent of bacterial suspensions such as dynamic clustering in semi-dilute suspensions \cite{karani2019}.

Here, we employ dense suspensions of Quincke random walkers to gain insights into the turbulent-like flows of bacterial and active fluids. We consider suspensions confined between two parallel walls with spacing comparable to the colloid diameter,   
and  experimentally measure spatial velocity correlations, structure functions, and kinetic energy spectra under varying concentrations of suspensions and activity, which is modulated by the applied field strength.

\section{Experiment}

Spherical polymethyl methacrylate (PMMA) spheres (Phosphorex) with a diameter of \( d = 100 \, \mathrm{\mu m} \) are dispersed in a \( 0.15 \, \mathrm{mol \, L}^{-1} \) AOT-hexadecane solution (conductivity $\sigma=2.2\times 10^{-8}\,\mathrm{S\cdot m}$). 
The experimental chamber consists of two  $7.5 \times 5$ cm$^2$ ITO coated glass slides (Delta Technologies), separated by a ring-shaped Teflon tape with thickness of 120 \SI{}{\micro\meter}, see Figure \ref{fig:setup}(a). The chamber is filled with the suspension solution.   
The particle motion and tracking is visualized using an inverted optical microscope (Zeiss) with 2X magnification mounted on a vibration isolation table (Kinetic Systems, Inc.). Videos were recorded at frame rates 500 or 1000 frames per second by a high speed camera (Photron SA 1.1). The system is powered by a pre-designed pulsed electric field supplied by a high voltage amplifier (Matsusada) controlled by a function generator (Agilent Technologies 33521A).
Particle tracking, flow field reconstruction and data analysis are carried out by means of particle image velocimetry (PIV) or particle tracking velocimetry (PTV) using open source Python codes (OpenPIV~\cite{alex2021} and Trackpy~\cite{allan2024}) as well as custom codes (see Appendix for details).

 When a constant voltage is applied across the ITO electrodes, the particles become polarized due to the accumulation of free charges at their interfaces, generating an induced electric dipole oriented antiparallel to the applied electric field. Beyond a critical field strength, the dipole loses its symmetric orientation, triggering what is known as the Quincke electrorotation \cite{Lemaire:2002, Jones:1984} and a particle  rolls along the surface with a constant speed \cite{bricard2013,pradillo2019}, see Figure \ref{fig:setup}(b).
  In our system, the typical threshold voltage for the onset of electrorotation is \( E_q = 1.6 \, \mathrm{V/\mu m} \) 
 and roller velocities  range from \( 4 \, \mathrm{mm/s} \) to \( 35 \, \mathrm{mm/s} \).

When the electric field is pulsating, as shown in Figure \ref{fig:setup}(c), the behavior of the roller changes. The roller ``runs'' while the field is on and stops when the field is turned off. Once the field is turned on again, the roller chooses a new random direction of rolling \cite{karani2019}, which results in diffusive motion of the colloid. 
The colloid run-stop-turn random walk mimics the bacterial run-and-tumble motility. A dense suspension of Quincke random walkers exhibits turbulent-like flow, see Figure \ref{fig:setup}(d)-(f). In our experiments, we vary the field magnitude \( E \) while maintaining fixed pulse durations \( \tau_{\mathrm{on}} = 40 \, \mathrm{ms} \) and intervals \( \tau_{\mathrm{off}} = 20 \, \mathrm{ms} \). The off-time is much longer than the particle memory time, which is the Maxwell-Wagner polarization relaxation time \( t_{\mathrm{MW}} = 2 \, \mathrm{ms} \), to ensure that the roller executes an uncorrelated random walk. \( \tau_{\mathrm{on}} \) sets the particle persistence length and it is chosen so that 
it is comparable to the collision time, which varies from \( 7 \, \mathrm{ms} \) to \( 38 \, \mathrm{ms} \). 
{The average collision time of the particles is estimated based on the average particle number density, $\tau_{\mathrm{col}} = \frac{1}{nd\langle |\mathbf{v}|\rangle}$, where $n$ is the number density of particles, $d$ is the particle's diameter, $\langle |\mathbf{v}|\rangle$ is the average speed of the particles.}

We use a ring-shaped Teflon tape as both a separator and a solid boundary to enclose the suspension solution. To prevent particles from becoming trapped between the glass slides and the tape during chamber assembly, the particles are concentrated to the central region of the chamber; 
the chamber is first filled with oil, and then a drop of concentrated suspension ($\sim 40 \mathrm{\mu L}$) is deposited in the middle of the chamber. 
The uniform particle region has a diameter of approximately 1.6 cm, which is twice the diameter of the recorded field of view. The total suspension region spans 2.2 cm in diameter. Due to the presence of a particle-free zone between the uniform particle region and the tape boundary, we observed a gradual decrease in suspension concentration within the field of view during the experiment. For the results reported later, the concentration value represents the average concentration in the field of view, with variations of less than $5\%$ over time.

\begin{figure}[h]
	\includegraphics{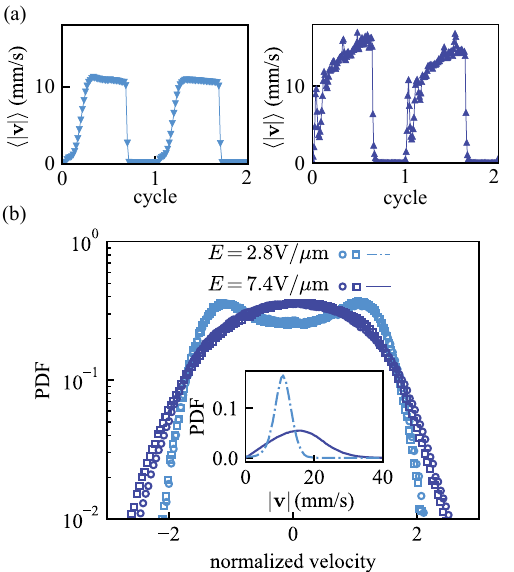}
	\caption{ \footnotesize {
(a) Flow development during the electric field pulsation: mean flow velocity as a function time. 
(b) Velocity distribution at \( E = 1.7E_q, \phi = 0.27 \) and  \( E = 4.6 E_q, \phi = 0.66 \). 
Inset: Flow speed follows a Boltzmann distribution.
}}
	\label{fig:general}
\end{figure}

\section{Results}
\subsection{Flow structure}

A unique feature of the Quincke random walkers is that when the field is on, they all run, and when the field is turned off, they all stop, see Fig. \ref{fig:general}(a). At lower activity and particle fraction,  \( E = 1.7 E_q, \phi = 0.27 \), the flow driven by the colloids reaches a  steady state within one cycle while the field is on. However, higher activity and particle numbers, \( E = 4.6 E_q, \phi = 0.66 \), give rise to a more vigorous flow, which does not reach steady state within the period of the applied electric field.  
The collective dynamics mirrors the behavior of the isolated Quincke random walker, see Appendix, Figure 6. The difference in the flow development is due to the particle inertia \cite{pradillo2019}. Fig. \ref{fig:general}(b) shows the velocity distribution for the experiments with low activity, low density (\( E = 1.7E_q, \phi = 0.2 \)) and higher activity, high density (\( E = 4.6 E_q, \phi = 0.6 \)) systems. A bimodal structure of the PDF suggests that the particles are running with similar velocity; the sampling of the random velocity directions results in the two peaks \cite{kokot2019diffusive}. 
At higher concentration, {the PDF approaches a Gaussian distribution with a single peak due to the fact that particles experience multiple collisions in the same time period between two consecutive frames, $t_{\mathrm{ fps}}$, used to determine the velocity ($\tau_{\mathrm{col}} < t_{\mathrm{ fps}}$).

\begin{figure}
	\includegraphics[width=0.75\textwidth]{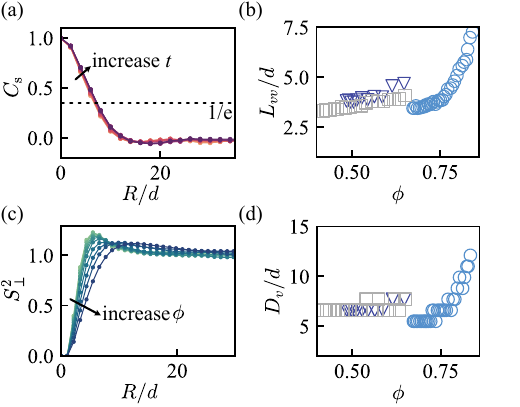}
	\caption{ \footnotesize {
(a) Velocity spatial correlation function, $C_{\mathrm{s}}(R)$, within one period. \( E = 1.7E_q, \phi = 0.84 \). The correlation length, $L_{vv}$, is defined at the location where $C_s(R) = 1/\mathrm{e}$. 
(b) Velocity correlation length $L_{vv}$ as a function of concentration ($\phi$) at  different field strengths (circle: $E = 1.7 E_q$; square: $E = 2.8 E_q $; triangle: $E = 4.6 E_q $). 
(c) Angular-averaged normalized two-point velocity correlation function vs normalized radial distance
at $E =1.7 E_q$ with {increasing $\phi$ from  0.67 to 0.85}. The peak position of $S_{\perp}^{2}$ indicates the effective vortex diameter $D_v$. 
(d) The effective vortex diameter increases with $\phi$.
}}
	\label{fig:cs}
\end{figure}

\begin{figure*}
	\includegraphics[width=\textwidth]{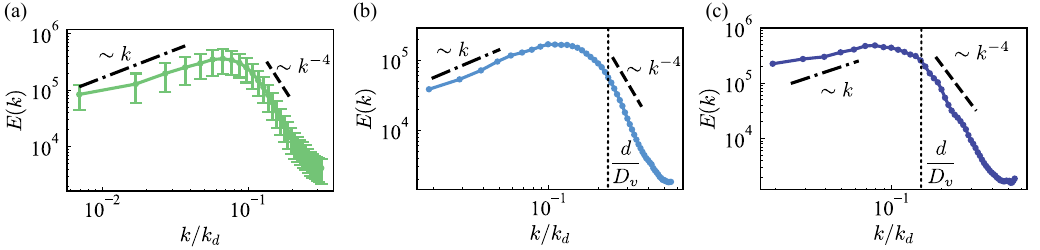}
	\caption{ \footnotesize {
(a) The kinetic energy spectrum exhibits similar behavior across varied concentrations and field strengths. 
\( E(k) \sim k \) at small \( k \) and \( E(k) \sim k^{-4} \) at large \( k \). The experiment is performed at \( E = 1.7 E_q, \phi = 0.07 \).
(b) The energy spectra at two extreme situations (left: \( E = 1.7E_q,\, \phi = 0.16 \), right: \( E = 4.6 E_q,\, \phi = 0.58 \)) exhibit the \( k^{-4} \) scaling at large \( k \), corresponding to lengthscales 
smaller the mean vortex size \( D_v \).
}}
	\label{fig:Ek}
\end{figure*}

To characterize the flow structure, we compute the  velocity spatial correlation function, VCF,
\begin{equation}
	C_{\mathrm{s}}(R) = \frac{\left\langle \mathbf{v}(r)\mathbf{v}(r+R)\right\rangle_r}{\left\langle\mathbf{v}^2(r)\right\rangle_r}
\end{equation}
where 
{$\mathbf{v}$ is the field velocity, and $\langle\cdot\rangle_r$ is the spatial average over all positions $r$.} 
The velocity field is obtained directly from Particle Image Velocimetry (PIV) for experiments with dense suspensions $\phi>0.5$ or by interpolation based on individual particle velocity result from Particle Tracking Velocimetry (PTV) for experiments with low particle density $\phi< 0.4$.  The VCF within the duration of one period of the applied field is plotted in Fig.~\ref{fig:cs}(a). It shows velocity anticorrelation (negative VCF) typical for presence of vortices. If $\tau_{\mathrm{on}}$ is sufficiently long, a single giant vortex---whose size is constrained by the system boundary--- forms, as previously observed in experiments under weaker confinement \cite{zhang2022a, bricard2013}. A single vortex would exhibit a velocity correlation value of $-1$. However, in our system, the presence of multiple vortices at different locations and  of varying sizes results in 
a shallower minimum.
Characteristic velocity correlation length, $L_{vv}$, defined from the decay of the velocity space correlation function, $C_s(L_{vv}) = 1/\mathrm{e}$, increases with the particle fraction, see Fig.~\ref{fig:cs}(b).

The velocity structure function provides further insight into the physics of active turbulence \cite{wensink2012, wei2024}.
{The normalized perpendicular structure function $S^2_{\perp}(R)$ is defined as:
\begin{equation}
	S^2_{\perp}(R) = \frac{\left\langle \left[ \left(\mathbf{v}(\mathbf{r} + \mathbf{R}) - \mathbf{v}(\mathbf{r})\right) \cdot \mathbf{e}_{\perp} \right]^2 \right\rangle_r}{\left\langle \mathbf{v}^2(\mathbf{r}) \right\rangle_r}
\end{equation}
}
Here, $\mathbf{e}_{\perp}$ is an  orthogonal directions to $\mathbf{R}$.
The maximum of $S^2_{\perp}$ corresponds to the effective vortex size $D_v$. Fig.\ref{fig:cs}(c) shows the $S^2_{\perp}$ as a function of $\phi$ 
at $E = 1.6 E_q$. The peak location, which corresponds to the mean vortex size $D_v$,  shifts 
 to larger $R$ with the density. 
 Likewise, the mean vortex size $D_v$ follows a similar trend as $L_{vv}$ with the particle number density $\phi$, see Fig. \ref{fig:cs}(d).

\subsection{Kinetic energy spectrum}

We next examine how the kinetic energy \( E \) is distributed across different length scales for varying field strengths and particle concentrations.
The energy spectrum, i.e., the energy of a vortex with size $\sim 1/k$, \( E(k) \) is computed  from  the velocity field \( \mathbf{v}(\mathbf{r}, t) \) through the relation
\cite{kraichnan1980, boffetta2012, martinez-prat2021, wensink2012, frisch1995} 
\begin{equation}
	E(k) = \frac{k}{4\pi L^2} \langle \vert \hat{\mathbf{v}}(\mathbf{k}) \vert^2\rangle
\end{equation}
where \( k = |\mathbf{k}| \) is the wavenumber, $L^2$ is the system area and $\hat{\mathbf{v}}(\mathbf{k}) = \int_{\mathbb{R}^2} \mathbf{v}(\mathbf{r}) e^{-i \mathbf{k} \cdot \mathbf{r}} \, d\mathbf{r}$ is the Fourier transformation of velocity. $E = L^2 \int_0^{\infty} E(k) dk$ is the total energy per unit mass of the system.  Fig. \ref{fig:Ek} shows the energy spectra for our system.

Different scaling laws have been predicted and observed for active nematics and bacterial systems, as reported in references \cite{giomi2015, martinez-prat2021, wei2024}. Despite the wide range of particle activities and concentrations in our experiments, the energy scaling law predominantly follows $E(k) \sim k^{-4}$ at large $k$. This scaling holds in a regime of wavenumber  smaller than the typical size of vortices, and when dissipation is dominated by  the flow,  rather than friction with the boundaries.

Our experiments also find exponential vortex size distribution (see Appendix Figure 7), which has been observed in active nematic turbulence and bacterial turbulence \cite{giomi2015, martinez-prat2021, wei2024}, where $E(k) \sim k^{-4}$ is found. This exponential distribution underlies the scaling for the energy density spectrum in such systems. The measurement of the vortex size $D_v$ from the previous section is indicated in the energy spectrum {Fig.\ref{fig:Ek}(b-c), and  roughly coincides with the wavenumber above which the \( k^{-4} \) scaling exists.

\subsection{Temporal memory effects}
In was recently demonstrated ~\cite{zhang2022natphys}  that Quincke rollers in a globally correlated collective state (vortex) spontaneously develop structural positional ordering that ''imprint'' the chiral state of the system. Under a cessation and restoration of the activity facilitated by a pulsed field with both $\tau_{\mathrm{on} }$ and $\tau_{\mathrm{off} }$ significantly larger than all relevant timescales,
 a global vortex alternates its chiral state with each activity cycle. This periodic response, despite the absence of memory at the level of individual particles, was attributed to a phenomenon referred to as collective or "structural memory".
To investigate whether a similar "memory" phenomenon occurs in our system under a pulsating field with shorter $\tau_{\mathrm{on}}$, we compute 
the temporal correlations function of the velocity field 
\begin{equation}
	C_{\mathrm{T,vf}}(\tau) = N^{-1} \sum_i {\bigl< \mathbf{v}_i(0)\cdot \mathbf{v}_i(\tau)} \bigr>_t / \bigl< \mathbf{v}_i^2(0) \bigr>_t
\end{equation}
and the vorticity field,
\begin{equation}
	C_{\mathrm{T,\omega}}(\tau) = N^{-1} \sum_i {\bigl< \omega_i(0)\cdot \omega_i(\tau)} \bigr>_t / \bigl< \omega_i^2(0) \bigr>_t
\end{equation}
{Here, the system is divided into a square grid.
$\mathbf{v}_i$ is the average velocity at the cell $i$; $N$ is the total number of cells and $\langle\cdot\rangle_t$ denotes the time average. Similarly, we compute the vorticity field $\omega_i$ based on the velocity field and proceed to compute the vorticity temporal correlation function. }

 \begin{figure}[!ht]
	\includegraphics{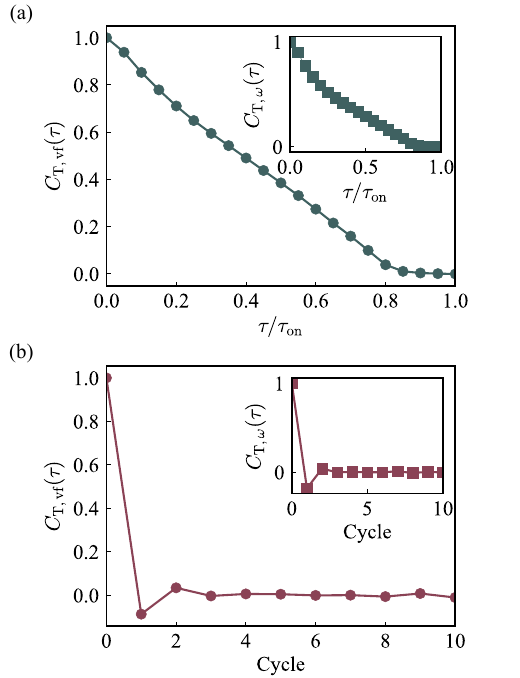}
	\caption{ \footnotesize {
(a) The temporal correlations of the velocity and vorticity fields within one period of application of the field. 
(b) The temporal correlations of the velocity and vorticity exhibit anti-correlation between subsequent field pulses. The finding indicates that at the subsequent cycle, a vortex is likely to occur at the same location but with opposite direction of rotation.
}}
	\label{fig:memory}
\end{figure}

Within one cycle, see Figure \ref{fig:memory}(a), the time correlation function decays to zero, indicating that at a fixed location, the flow field  decorrelates from its initial state, likely due to the particle collisions. Nevertheless, when the time correlation function is computed between the pulses (stroboscopically- at fixed time within each period), the correlation functions exhibit clear non-negligible anti-correlations after one cycle (see Figure \ref{fig:memory}(b)) indicative  of the velocity and vorticity reversals in the ensemble during temporal activity cycling. 
This behavior is attributed to structural memory, facilitated by positional ordering of particles in self-organized vortices\cite{zhang2022natphys}. 
In our system under short activity pulses, local vortices do not coalesce into a giant vortex. %
Nonetheless, the formation of these locally correlated vortical states is the source of the anti-correlations in the system between the activity cycles. Only particles participating in local vortices contribute to the anti-correlations resulting in the correlation value significantly lower that \(-1\) (as it is with a global vortex state~\cite{zhang2022natphys}). The signature of the collective ensemble memory is robust and easily detectable. Over time (beyond 2 cycles) the correlations between the activity cycles are suppressed as the locations of vortices get randomized.  

\section{Conclusion}

We have investigated active turbulence in a system of Quincke random walkers, a synthetic system which mimic the run-and-tumble behavior of bacteria. The Quincke random walkers provide a versatile experimental model of  active matter to study the fundamental mechanisms of active turbulence.
We demonstrated that the kinetic energy spectra follow  \(k^{-4}\) scaling, consistent with the observations in very confined bacterial suspensions \cite{wei2024}. 
Our experiments indicate exponential vortex size distribution, which has been previously observed in active nematic turbulence and bacterial turbulence. 
In addition, we have found clear evidence of structural collective memory in this quasi-two-dimensional dense suspension.
Our findings highlight the importance of transient dynamic memory in the ensemble, both at the particle level and collective, in accessing unconventional self-organized states. 
\section{Acknowledgment}

The research of R.L. and P.V. was supported by NSF-DMR award 2004926. The research of A.S. was supported by the U.S. Department of Energy, Office of Science, Basic Energy Sciences, Materials Sciences and Engineering Division.

\appendix

\section{Isolated Quincke Random Walker}
The speed of an isolated Quincke random walker takes time to reach a steady state. At $E = 1.7E_q$, the speed eventually stabilizes to a steady state. However, at $E = 4.6E_q$, the particle's speed exhibits oscillations with a larger amplitude when the field is applied, and it continues to increase throughout the runtime $\tau_{\mathrm{on}}$. 

\begin{figure}[!ht]
	\includegraphics{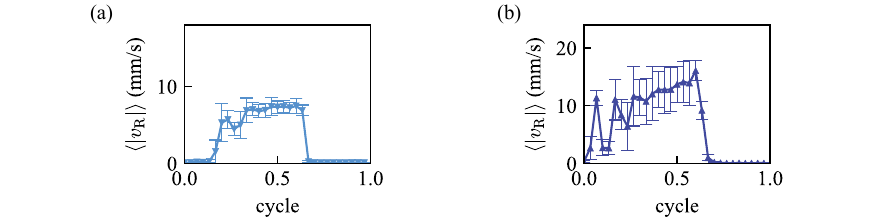}
	\caption{Speed measurements of an isolated Quincke random walker at different field strengths: (a) $E = 1.7E_q$ and (b) $E = 4.6E_q $. The data represent averages over 10 cycles for an isolated particle. Error bars indicate the standard deviation of the measured speed.}
\end{figure}

\section{Experimental Imaging and Analysis Methods}

\subsection{Imaging Setup and Parameters}
For imaging, a high-speed camera (Photron SA 1.1) is employed in conjunction with a microscopy setup (Zeiss, 2x magnification). The frame rate (fps) is strategically selected to be either 500 or 1000, depending on the particle concentration and the applied field strength. This selection ensures that the motion of the particles is adequately resolved, facilitating accurate tracking and analysis of their dynamics under the experimental conditions.

\subsection{Velocity and Vorticity Measurements}

\subsubsection{Dilute Suspensions: Particle Tracking Velocimetry (PTV)}
In dilute suspensions (area fraction $\phi<0.4$), Particle Tracking Velocimetry (PTV) is employed to measure individual particle velocities. The Trackpy\cite{allan2024} library is utilized for this purpose. When dealing with images captured at different frame rates (fps) under varying field strengths or concentrations, the time interval used to link particles is carefully chosen such that the displacement of particles between frames remains within $1/4$ to $1/2$ of the particle diameter. This ensures accurate tracking without excessive ambiguity or loss of particles due to large displacements. The flow field is then reconstructed using cubic interpolation of the scattered particle data points to generate interpolated velocity values on a predefined grid. This interpolation is facilitated by the \texttt{scipy.interpolate.griddata} function with the \texttt{cubic(2D)} method.

\subsubsection{Dense Suspensions: Particle Image Velocimetry (PIV)}
For dense suspensions with an area fraction of particles $\phi \ge 0.5$, the suspension is treated as a continuum phase. In this regime, Particle Image Velocimetry (PIV) is utilized to determine the velocity field. The OpenPIV\cite{alex2021} library in Python is employed to estimate the 2D velocity field on a 2D cubic grid. Averaging windows of size 20 pixels by 20 pixels are selected, corresponding approximately to 2 by 2 times the mean particle diameter of $100 \, \mathrm{\mu m}$ for Polymethyl methacrylate (PMMA) particles. This window size ensures statistical reliability while resolving spatial flow field structures on the order of a few bacterial lengths. A 50\% overlap between neighboring bins is used to enhance spatial resolution and accuracy.

\subsubsection{Vorticity Calculation}
The vorticity of the flow is computed using a second-order central difference scheme, implemented through the \texttt{numpy.gradient} function, which calculates the spatial derivatives required for vorticity determination. The results were cross-validated using Richardson's Extrapolation (four-point central), confirming consistency.

\subsubsection{Vortex size distribution based on Okubo-Weiss
(OW) parameter}
The flows generated by our active system exhibit distinct vortex structures.
In the main text, we employ the structure function to compute the average vortex size. Here, we further identify and characterize the size distribution of vortices using the Okubo-Weiss (OW) parameter \cite{giomi2015,martinez-prat2021}. The OW parameter is defined as
$\text{OW} = (\partial_x v_x + \partial_y v_y)^2 - 4(\partial_x v_x)(\partial_y v_y) + 4(\partial_x v_y)(\partial_y v_x).$

Regions where $\mathrm{OW} < 0$ are identified as vortex regions. To analyze the vortex size distribution, we compute the histogram of the areas of connected vortex regions. This distribution, denoted as $n(a) = N(a)/\sum_a N(a)$, represents the normalized frequency of vortices with area $a$, where $N(a)$ is the number of vortices of area $a$. We find that $n(a)$ follows an exponential distribution, $n(a) \propto \exp(-a/a^*)$, where $a^*$ is the mean vortex area and $D_{\mathrm{OW}}=2\sqrt{a^*/\pi}$ is the mean vortex diameter. In Figure \ref{fig:ow} we show the distribution of vortex size for two typical experiments reported in the main text. Their fitted mean vortex diameters are $0.43\,\mathrm{mm}$ and $0.61\, \mathrm{mm}$, which are comparable from the mean vortex diameter measured from structure function in the main text.

The exponential distribution of vortex areas is a key assumption in the theoretical framework used to explain the scaling of the energy spectrum in active systems, such as bacterial suspensions \cite{wei2024} and active nematics \cite{giomi2015,martinez-prat2021}.

\begin{figure}[!ht]
	\includegraphics{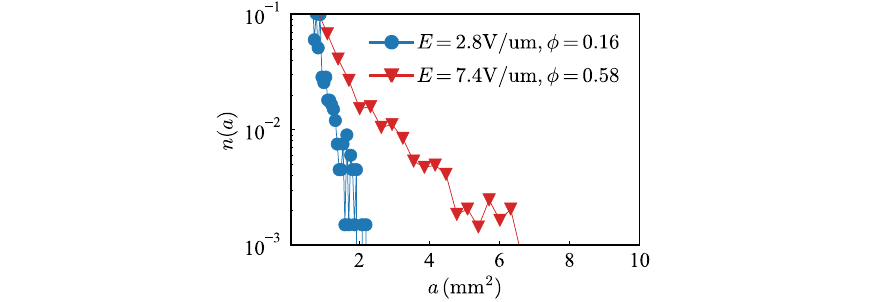}
	\caption{Vortex size distribution obtained from two representative experiments. The distributions exhibit an exponential decay.}
	\label{fig:ow}
\end{figure}
\subsection{Kinetic Energy Spectra}
\label{sec:energy_spectra}

\subsubsection{Energy Spectra Calculation}
\label{sec:es}

The total kinetic energy (per unit mass) of a 2D system is defined in real space as:
\begin{equation}
E = \frac{1}{2} \int_{\mathbb{R}^2} \mathbf{v}(\mathbf{r}) \cdot \mathbf{v}(\mathbf{r}) \, d\mathbf{r}.
\end{equation}

To analyze spectral energy distribution, we express the velocity field through its Fourier transform,
\begin{equation}
\mathbf{v}(\mathbf{r}) = \frac{1}{(2\pi)^2} \int_{\mathbf{k}} \hat{\mathbf{v}}(\mathbf{k}) e^{i \mathbf{k} \cdot \mathbf{r}} \, d\mathbf{k},
\end{equation}
where the Fourier coefficients are given by,
\begin{equation}
\hat{\mathbf{v}}(\mathbf{k}) = \int_{\mathbb{R}^2} \mathbf{v}(\mathbf{r}) e^{-i \mathbf{k} \cdot \mathbf{r}} \, d\mathbf{r}.
\end{equation}

Relating  
\begin{equation}
\begin{aligned}
\int_{\mathbf{k}} \hat{\mathbf{v}}(\mathbf{k}) \cdot \hat{\mathbf{v}}^*(\mathbf{k}) \, d\mathbf{k} 
&= \int_{\mathbf{k}} \int_{\mathbb{R}^2} \int_{\mathbb{R}^2} \mathbf{v}(\mathbf{r}) \cdot \mathbf{v}(\mathbf{r'}) e^{-i \mathbf{k} \cdot (\mathbf{r'} - \mathbf{r})} \, d\mathbf{r} \, d\mathbf{r'} \, d\mathbf{k} \\
&= \int_{\mathbb{R}^2} \int_{\mathbb{R}^2} \mathbf{v}(\mathbf{r}) \cdot \mathbf{v}(\mathbf{r'}) \int_{\mathbf{k}} e^{-i \mathbf{k} \cdot (\mathbf{r'} - \mathbf{r})} \, d\mathbf{k} \, d\mathbf{r} \, d\mathbf{r'} \\
&= (2\pi)^2 \int_{\mathbb{R}^2} \int_{\mathbb{R}^2} \mathbf{v}(\mathbf{r}) \cdot \mathbf{v}(\mathbf{r'}) \delta(\mathbf{r'} - \mathbf{r}) \, d\mathbf{r} \, d\mathbf{r'} \\
&= (2\pi)^2 \int_{\mathbb{R}^2} \mathbf{v}(\mathbf{r}) \cdot \mathbf{v}(\mathbf{r}) \, d\mathbf{r},
\end{aligned}
\end{equation}
using the orthogonality of Fourier modes,
\begin{equation}
\int_{\mathbf{k}} e^{-i \mathbf{k} \cdot (\mathbf{r'} - \mathbf{r})} \, d\mathbf{k} = (2\pi)^2 \delta(\mathbf{r'} - \mathbf{r}),
\end{equation}
we simplify to,
\begin{equation}
E = \frac{1}{2(2\pi)^2} \int_{\mathbf{k}} |\hat{\mathbf{v}}(\mathbf{k})|^2 d\mathbf{k}.
\end{equation}

To derive the radial energy spectrum $E(k)$, we convert to polar coordinates $(k, \phi)$ and average over angular dependence:
\begin{equation}
E = \frac{1}{2(2\pi)^2} \int_0^\infty k \left(\int_0^{2\pi} |\hat{\mathbf{v}}(k, \phi)|^2 d\phi\right) dk.
\end{equation}

For homogeneous isotropic turbulence, the energy depends only on wavenumber magnitude $k$. Defining the angle-averaged power spectrum,
\begin{equation}
\langle |\hat{\mathbf{v}}(k)|^2 \rangle_\phi = \frac{1}{2\pi} \int_0^{2\pi} |\hat{\mathbf{v}}(k, \phi)|^2 d\phi,
\end{equation}
we obtain the energy spectrum,
\begin{equation}
E(k) = \frac{k}{4\pi L^2} \langle |\hat{\mathbf{v}}(k)|^2 \rangle_\phi,
\label{eq:Ek_direct}
\end{equation}
which satisfies $E = L^2\int_0^\infty E(k)dk$ and $L^2$ is the area.

\subsubsection{Wiener-Khinchin Theorem}
\label{sec:wiener}

The energy spectrum can alternatively be computed from the velocity correlation function $C_{vv}(\mathbf{R}) = \langle \mathbf{v}(\mathbf{r}) \cdot \mathbf{v}(\mathbf{r} + \mathbf{R}) \rangle_{\mathbf{r}}$ via the Wiener-Khinchin theorem\cite{frisch1995},
\begin{equation}
\frac{|\hat{\mathbf{v}}(\mathbf{k})|^2}{(2\pi)^2} = \int C_{vv}(\mathbf{R}) e^{-i \mathbf{k} \cdot \mathbf{R}} \, d\mathbf{R}.
\end{equation}

Substituting into the expression for $E(k)$,
\begin{equation}
E(k) = \frac{\pi}{A} k \left\langle \int C_{vv}(\mathbf{R}) e^{-i \mathbf{k} \cdot \mathbf{R}} \, d\mathbf{R} \right\rangle_{\phi},
\label{eq:Ek_correlation}
\end{equation}
where $\langle \cdot \rangle_{\phi}$ indicates averaging over the direction of $\mathbf{k}$ (azimuthal angle $\phi$).

\subsubsection{FFT and Windowing Effect}
The derivation in \ref{sec:wiener} used Wiener-Khinchin theorem to relate the energy spectrum with velocity spatial correlation function, which is valid in infinite domain. Our experiment is neither infinite nor periodic. The discontinuity of the boundary could effect the energy spectrum. We use a Hanning window\cite{wilken2020} to process the velocity field measured from experiment, before computing energy spectrum directly (\eqref{eq:Ek_direct}) or from  correlation function (\eqref{eq:Ek_correlation}). The Hanning window is defines as $w(x,y) = \frac{1}{4} \left[1-\cos(2\pi x/L)\right]\left[1-\cos(2\pi y/L)\right]$. $L$ is the size of the square region of velocity field. We would like to point out the result difference from no windowing procedure and correlation equation (\eqref{eq:Ek_correlation}), which shows the error of energy spectrum result without dealing with the in-continuous and non-periodic experimental data. In the main test, we use the direct method (\eqref{eq:Ek_direct}).
\begin{figure}[!ht]
	\includegraphics{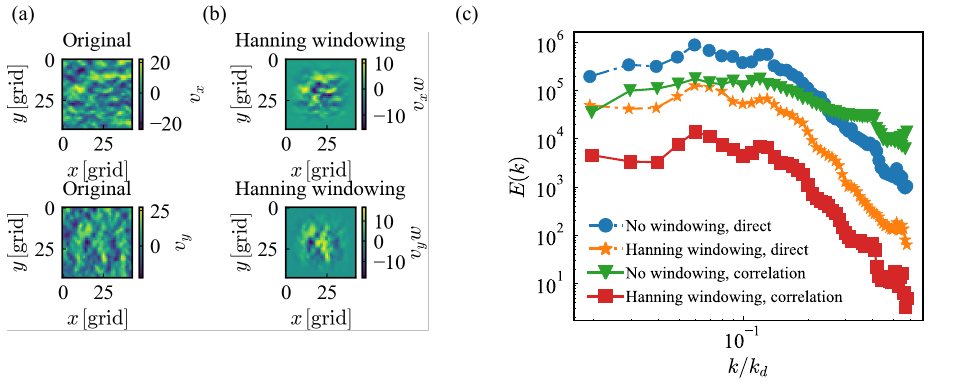}
	\caption{(a) Original velocity field based on PIV. (b) Velocity field after applying Hanning window. (c) Comparison of the energy spectrum based on direct method or correlation function.}
\end{figure}



\bibliographystyle{unsrt}

\end{document}